\documentclass[%
 reprint,aps, prl,  showpacs,
 amsmath,amssymb,
]{revtex4-1}
\usepackage{graphicx}
\usepackage{dcolumn}
\usepackage{bm}
\usepackage[utf8]{inputenc}
\usepackage[T1]{fontenc}
\usepackage{lmodern}
\usepackage[english]{babel}
\usepackage{siunitx}
\usepackage{microtype}
\usepackage{textcomp}
\usepackage{amssymb}
\usepackage{upgreek}
\usepackage{epstopdf}

\begin{document}

\preprint{APS/123-QED}

\title{Motor-free actin bundle contractility driven by molecular crowding}

\author{Jörg Schnauß$^1$}
\author{Tom Golde$^1$}
\author{Carsten Schuldt$^1$}
\author{B. U. Sebastian Schmidt$^1$}
\author{Martin Glaser$^1$}
\author{Dan Strehle$^1$}
\author{Claus Heussinger$^2$}
\author{Josef A. Käs$^1$}
\affiliation{1 Institute for Experimental Physics I, University of Leipzig, Linnéstraße 5, 04103 Leipzig, Germany}
\affiliation{2 Institute for Theoretical Physics, Georg-August University of Göttingen, Friedrich-Hund Platz 1, 37077 Göttingen, Germany}
\date{\today}

\begin{abstract}
  Modeling approaches of suspended, rod-like particles and recent
  experimental data have shown that depletion forces display different
  signatures depending on the orientation of these particles. It has
  been shown that axial attraction of two rods yields
 contractile forces of \SI{0.1}{\pico\newton} that are independent of
  the relative axial shift of the two rods.
  Here, we measured depletion-caused interactions of actin bundles
  extending the phase space of single pairs of rods to a
  multi-particle system. In contrast to a filament pair, we found
  forces up to \SI{3}{\pico\newton}. 
  Upon bundle relaxation forces decayed exponentially with a mean
  decay time of \SI{3.4}{\second}. These different dynamics are
  explained within the frame of a mathematical model by taking
  pairwise interactions to a multi-filament scale. The macromolecular
  content employed for our experiments is well below the crowding of
  cells. Thus, we propose that arising forces can contribute to
  biological force generation without the need to convert chemical
  energy into mechanical work.
\end{abstract}

\pacs{87.15.H-, 87.16.Ln, 87.15.kr, 87.16.Ka}

\maketitle
Interactions of actin and its molecular motor myosin are known as the fundamental process for biological force generation. These interactions convert chemical energy into mechanical work by ATP hydrolysis \cite{Lodish.op.2000}. However, we show an alternative mechanism of force generation in the absence of any molecular motors or actin accessory proteins. The system is not driven by ATP hydrolysis and relies solely on minimization of the free energy based on filament - filament interactions. Interactions are induced by a crowded environment in a regime well below the macromolecular content of biological cells \cite{Ellis.2001}.\par
These so-called depletion forces were originally described by spherical colloidal particles suspended in a polymeric solution \cite{Asakura.1958, Asakura.1954}. However, this effect inherently appears in crowded solutions independent of the geometry of colloidal particles. Besides lateral particle attraction, the influence of depletion forces on axially shifted rod-like colloids has already been described by theoretical approaches \cite{Kinoshita.2004, Li.2005, Galanis.2010}. All these approaches describe the relative shift of two rod-like particles due to the induced interaction. Arising forces are found to be linear in the axial shift since the energy gain per unit length is constant.\par
Recently, Hilitski et al. experimentally verified these approaches by investigating the overlap of single microtubuli filaments \cite{Hilitski.2014}. They found a constant force driving these two rods towards a maximized overlap of their excluded volumes. Additionally, force components sum up in a pairwise manner when introducing a third rod to the system \cite{Hilitski.2014}. We describe a different, emerging behavior of rod-like colloids in a multi-filament system, in our case actin bundles formed by the depletant methyl cellulose \cite{Hosek.2004, Strehle.2011, LauA.W.C..2009, Streichfuss.2011}. These bundles are formed without influences of additional accessory proteins. \par
We used a mesoscopic approach allowing to deflect bundles from their energetic minimum by pulling forces exerted by optical tweezers. We investigated kinetics and restoring forces arising from the relative, axial sliding of single rod-like filaments within the bundle. Observed responses did not yield a constant force - in contrast to a two filament system - but an exponential force decay. These unexpected, complex dynamics can be explained by a mathematical model when increasing pairwise, linear interactions to a multi-filament scale. Additionally, the model is verified by simulations. These emergent dynamics can exert forces corresponding to a regime of weak active behavior of single myosin motors \cite{Carvalho.2013, Clemen.2005}.\par
\begin{figure}[t]
  \centering	
  \includegraphics[width=0.45\textwidth]{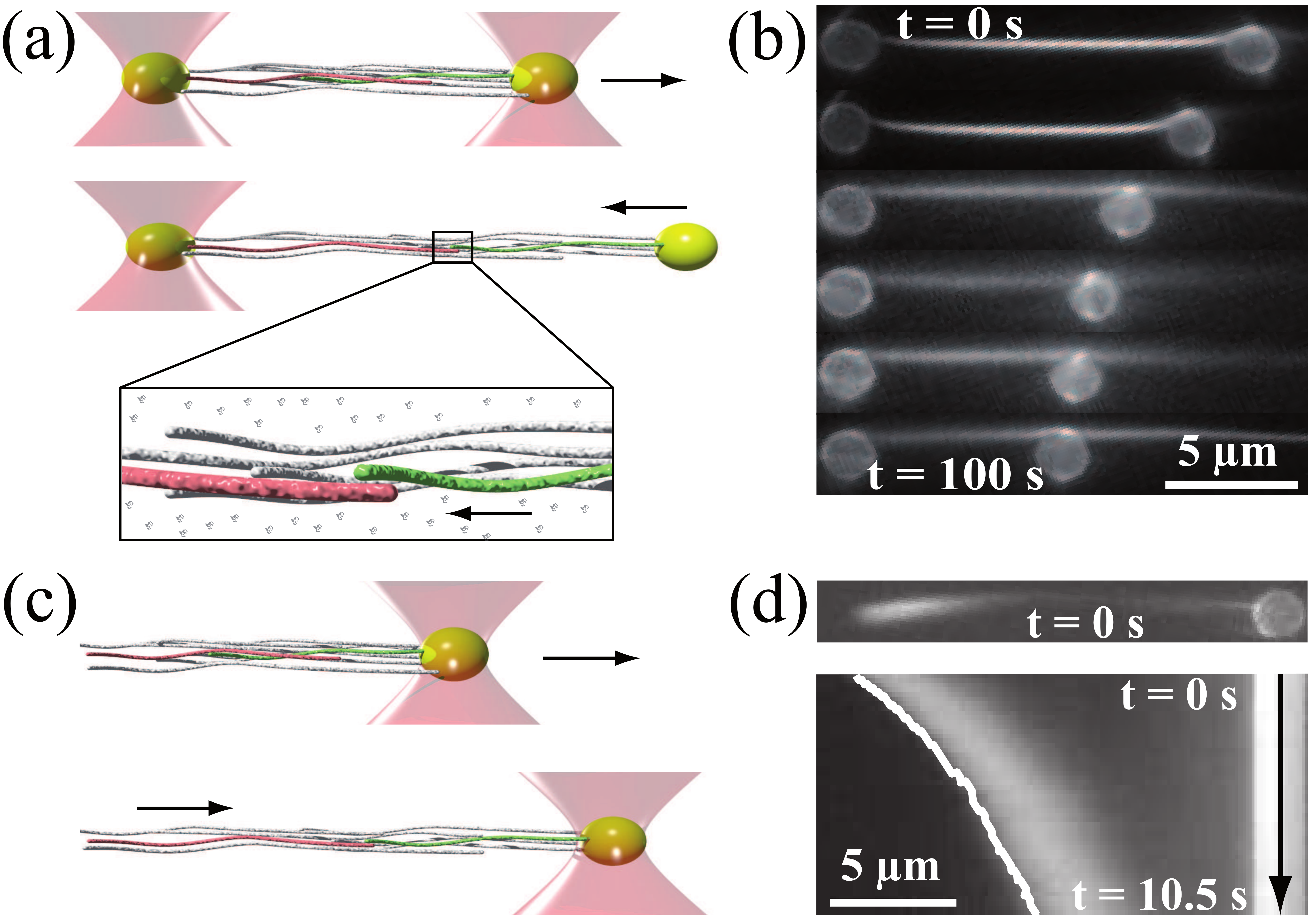}
  \caption{(Color online) (a) Optical tweezers were used to stretch bundles exceeding normal elastic deformations. After releasing one bead from the trap the bundle started to contract (Movie S 1). In a stretched bundle, the overlap of excluded volumes was not maximized anymore. When the pulling force was switched off, filaments tended to maximize this overlap again and contractions appeared (magnification). A bead tracking algorithm was used to transform recorded image series to bead trajectories giving the bundle length between these two beads over time. (b) After the pulling process the right bead was released and the bundle relaxed to a position maximizing the overlap of the excluded volumes again. (c) The bead can be trapped and moved through the viscous solution stretching the bundle due to friction. When the movement is stopped, the bundle started to contract. (d) Displayed are the first frame of the picture series and the kymograph of a contractile actin bundle attached to one bead with detected bundle length over time.}
	\label{fig:methods_overview}
	\end{figure}

To probe these contractions, we used two different experimental approaches. A dual-trap configuration was used to maneuver one bead while the other bead was held at a constant position. This pulling process resulted in a stretched bundle exceeding its former contour length. After releasing the deflected bead from the trap, the bundle started to contract (Fig. \ref{fig:methods_overview} (a)). This process was recorded as an image series of the fluorescent signals (Fig. \ref{fig:methods_overview} (b), Movie S 1). A bead tracking algorithm - computed with adapted MATLAB routines provided by Pelletier et al. \cite{Pelletier.2009} - was used to transform these image series to trajectories. Following data evaluations were conducted with self-written MATLAB scripts as described previously \cite{Golde.2013}. Alternatively, an arrangement of only one bead attached to a bundle was employed. By displacing the bead, the bundle was dragged through the solution. The longitudinal viscous drag elongated the bundle (Fig. \ref{fig:methods_overview} (c)). Immediately after the movement was stopped bundles started to contract and fluorescent signals were recorded (Movie S 2). To evaluate these experiments, a kymograph (picture series joined in one image) was used to visualize the bundle length over time (Fig. \ref{fig:methods_overview} (d)). Standard edge detections routines of MATLAB were employed to extract the bundle length at given times. \par
These methods are suitable to investigate dynamics of the system. However, multi-filament systems involve a variety of parameters (such as molecular content, bundle thickness, filament length distributions and more) yielding diverse starting conditions for every experiment.\par
In  our experiments we tested responses of filament bundles under
stress and recorded strains exceeding normal elastic deformations (up
to \SI{175}{\percent} of the initial contour length). Due to actin’s
rigidity these elongations can neither be attributed to thermal
fluctuations of single filaments nor stretching of the filament
backbone. Thus, within the pulling process filaments were pulled apart
and overlapping excluded volumes of filaments were not maximized
anymore. After stress release bundles started to contract. This
behavior can be attributed to filaments restoring a maximal overlap of their excluded volumes.\par
\begin{figure}
  \centering	
  \includegraphics[width=0.48\textwidth]{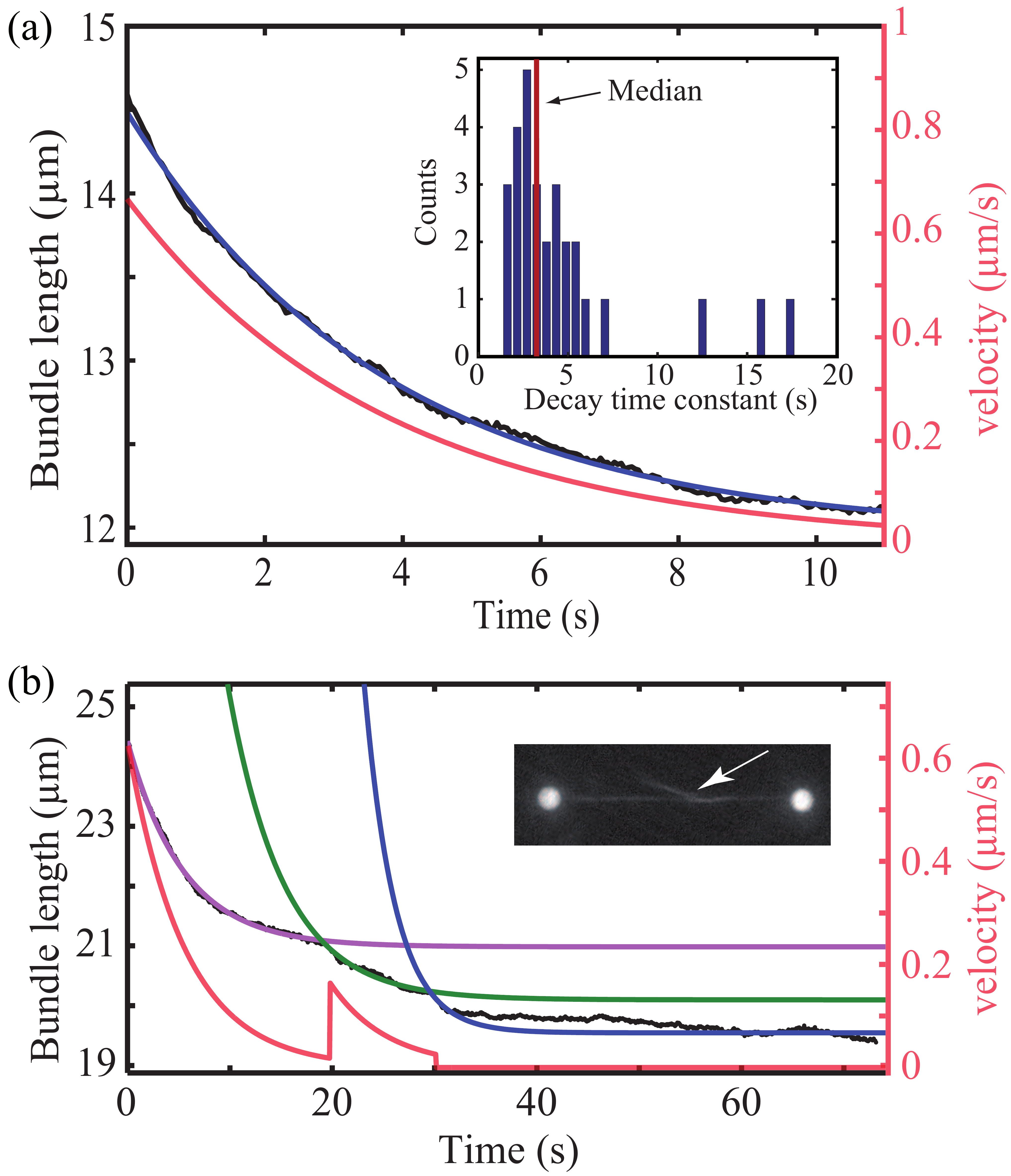}
  \caption{(Color online) (a) The recorded bundle length over time is
    well described by an exponential decay allowing an evaluation of
    the contraction velocity (red graph). Decay times (inset) are
    consistent with a median of \SI{3.4}{\second}. (b) A contracting
    bundle can exhibit multiple contraction events (due to a split
    bundle structure) described by a series of exponential decay
    functions with consistent decay times. The black curve displays
    the bundle contraction overlaid with according single exponential
    decay functions. The red graph is the according velocity for the
    contractions and accelerating events.}
  \label{fig:expfit}
\end{figure}

Although previous studies revealed a constant force for pairwise
overlapping filaments, the decreasing bundle length over time in our
experiments is well described by an exponential decay
(Fig. \ref{fig:expfit} (a)). Thus, bundle dynamics correspond to an overdamped relaxation in a harmonic free energy landscape.
These dynamics arise due to the
multi-filament nature of probed actin bundles as described in the
mathematical model below. Resulting exponential decay functions
($\mathrm{{bundle~length}(t) = a \cdot \mathrm{exp}(-t/\tau) + c}$)
yield a distribution of decay times $\tau$ showing the consistency of
the effect with a median of \SI{3.4}{\second} (Fig. \ref{fig:expfit}
(a)) inset). Determined exponential decays were used to calculate the
velocity of contractions showing maximal speeds in the range from 0.10
to \SI{0.65}{\micro\meter \per \second} (Fig. S2). Resulting maximal
forces were evaluated by Stokes’ law and typically range from 0.5 to \SI{3.0}{\pico\newton} (Fig. S2).\par
\begin{figure}

  \centering	
  \includegraphics[width=0.465\textwidth]{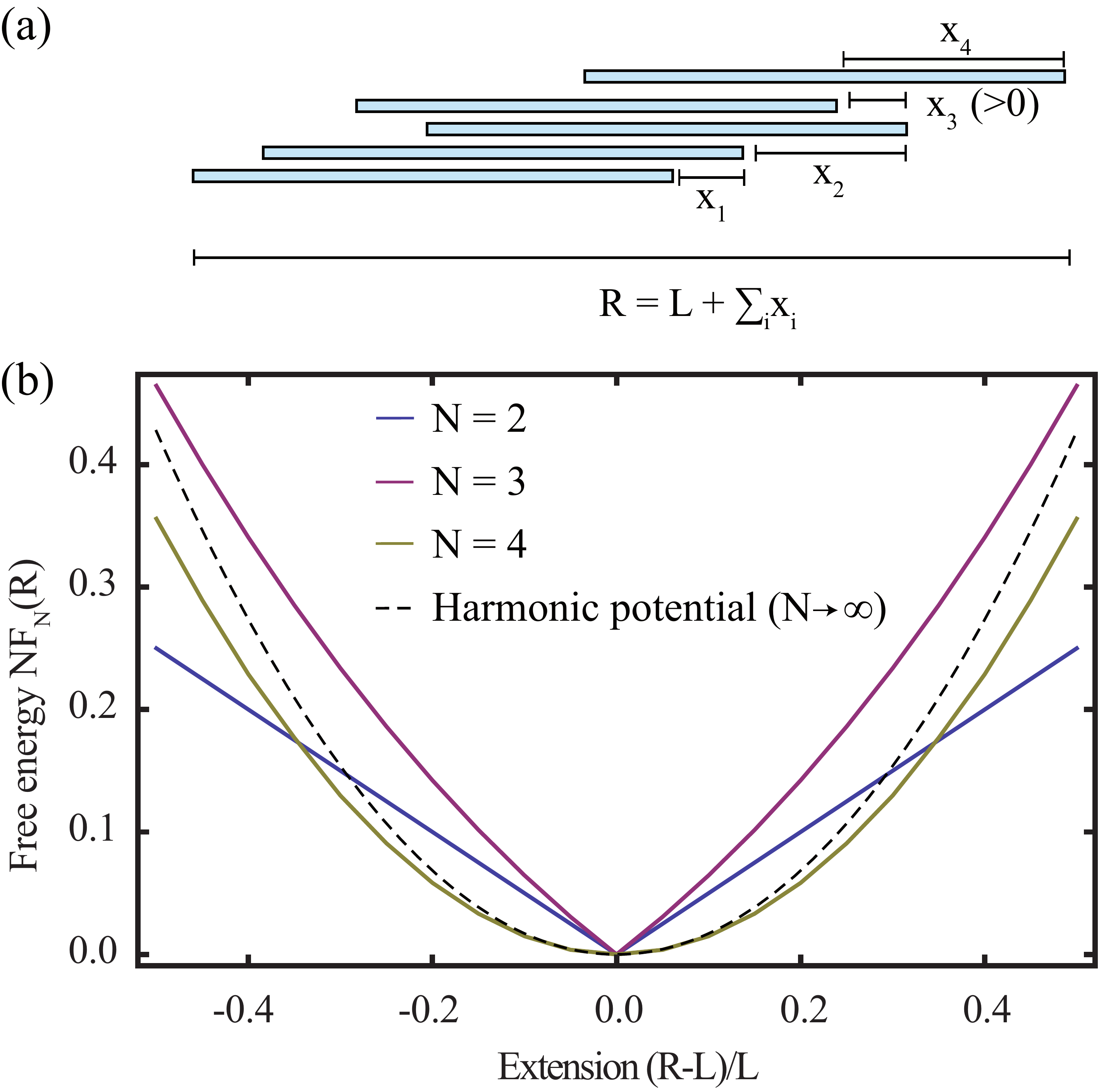}
  \caption{(Color online) (a) Schematic of idealized 2d-scenario, where forces are
    applied at the first (i = 1) and at the last filament (i = N). (b)
    Free energy $\mathrm{F_N (R)}$ vs. end-to-end distance R.  A
    two-filament bundle (N = 2) has a linear energy landscape, but
    with only a few filaments (N = 4) the asymptotic harmonic form
    (dashed) is nearly reached.}
  \label{fig:model}
\end{figure}

In three cases we observed contractions involving additional
accelerating events. These rendered one single exponential decay
inappropriate (Fig. 2 B) to describe the whole contraction
process. Those contractions, however, can be well described by a
series of exponential decay functions. Interestingly, the decay times
of these individual exponential decay functions are consistent. In
general, we attribute these accelerating events to split bundle
structures. A part of the bundle with originally overlapping filaments
was fully detached during the stretching process. Filaments in the
“main bundle” still shared excluded volumes to cause
contractions. When releasing the external stress these bundles started
to contract lacking the contribution of the non-overlapping
filaments. Their formerly attractive potential was not involved in the
cumulative energy balance of the starting contraction. At a certain
point non-overlapping filaments came close enough to share excluded
volumes with the already contracting bundle. New overlaps changed the
attractive potential and accordingly the energy balance. Thus, a
second or third additional internal contraction process set in and the
overall contraction was accelerated again (Fig. \ref{fig:expfit}
(b)).\\
To model the results of our experiments we extend the
depletion-induced interaction between filaments from individual
filament pairs to a multi-filament scale. Within the model a bundle is
represented as a two-dimensional arrangement of $\mathrm{N}$ rigid rods of
length $\mathrm{L}$. The only degree of freedom of the rods is their relative
axial shift $\mathrm{x_i}$ (Fig. \ref{fig:model}). The Hamiltonian is given by
\begin{equation}
	{\mathrm{H=-u\sum_{i=1}^{N-1}\left(L-|x_i|\right)-f \sum_{i=1}^{N-1}x_i,}}
\end{equation}
where the first term represents the depletion-induced attraction
between filaments; assumed pair-wise additive and of strength $\mathrm{u}$. The
second term is the work done by the external pulling force $\mathrm{f}$. The
free energy $\mathcal{F}$ and the force-extension relation
$\mathrm{\left\langle R-L \right\rangle = -\partial F / \partial f}$
can easily be calculated numerically. In the large-$\mathrm{N}$ limit one
obtains in linear response
\begin{equation}
  \mathrm{\left\langle R-L \right\rangle} = \mathrm{f\frac{N\left\langle x^2
      \right\rangle}{k_BT},} 
\end{equation}
which is a consequence of the law of large numbers (similar as, for
example, in the Gaussian chain model). The value $\mathrm{\left\langle
    x^2 \right\rangle}$ represents fluctuations
of a single filament pair in the absence of force (f=0). Neglecting
end-effects it is given by $\mathrm{\left\langle
    x^2 \right\rangle=2/(\beta u)^2}$.
As a result we find the force to be proportional to the extension with
a spring constant $\mathrm{k = u^2/N k_B T}$. In Fig. \ref{fig:model}
(a) we numerically calculate the free energy of this model as a
function of the bundle extension $\mathrm{R-L}$ and the number of
filaments $\mathrm{N}$ that are arranged laterally. With two filaments in the
arrangement (one pair), the free energy is a linear function of bundle
extension, as expected from the definition of the model. However, this
linear relation does not persist for multi-filament
arrangements. Already bundles with four filaments display
approximately a harmonic free energy that very closely resembles the
asymptotic form (N $\rightarrow \infty$).

Thus, within the analytical model a combination of several linear
force pairs in an additive manner yields a relation describing a
harmonic potential. The origin of this transition is the addition of
more and more internal degrees of freedom (in our case the relative
sliding of each individual pair) contributing an entropic term to the
free energy. Extended bundles have a much smaller entropy, because the
accessible configuration space for bundle conformations is highly
reduced.\par
While for the formulation of the theory several simplifying
assumptions have been made, we expect this scenario to be generic and
also useful to understand the complex experimental bundle
contraction. The internal degrees of freedom in this case may also
include filament bending fluctuations which are suppressed by
extension \cite{MacKintosh.1995}, thus decreasing the entropy. The
depletion force, in general, cannot be written as a sum over two-body
contributions \cite{Dijkstra.2002}. We test this assumption of the
model by running molecular dynamics simulations where the depletant is
modeled explicitly via soft spheres (see supplement). Simulations and
theoretical model are in excellent agreement indicating that many-body
effects for the depletion interaction in our case are indeed
negligible. Although other studies revealed a constant force for
pairwise overlapping filaments \cite{Kinoshita.2004, Li.2005,
  Galanis.2010}, we are able to show that pairwise linear, additive
forces create a harmonic potential. This potential for attractive
filament-filament interactions in our system supports the approach to
describe actin bundle contractions by an overdamped harmonic
motion. Within the frame of this model decreasing bundle lengths over
time can be well described by an exponential decay.\\
Our approach explains the spring-like elastic behavior of the bundle
under elongation predicting a spring constant $\mathrm{k=c \cdot
  u^2/(k_B T)}$. The prefactor $\mathrm{c \propto N_{3d}/N}$ can be
estimated to depend on the number of filament pairs N in the
two-dimensional bundle element ( Fig. \ref{fig:model} (a)) and the
number N$_{3d}$ of these elements that are coupled in parallel. The
precise value of c depends on the internal bundle structure and may
vary with the experimental situation. With c = O(1) we are able to
estimate a spring constant by using filament-filament interaction
energies of \SI{30}{k_B T \per \micro\meter} as measured previously
\cite{Streichfuss.2011}. The resulting k=\SI{3.6}{\pico\newton \per
  \micro\meter} is in good agreement with the magnitudes of our
data. We were not able to quantitatively compare this theoretical
approach to our data any further due to experimental
uncertainties. Unfortunately, there is no technique known to us to
determine the exact amount of filaments within the bundle \textit{in
  situ} and values can be only estimated roughly
\cite{Strehle.2011}. Furthermore, packing effects within the bundle
cannot be resolved, which would be essential to extend our model from
a simple two-dimensional arrangement to fully three dimensional
structures and to determine N$_{3d}$. Computer simulations are
underway to elucidate the specific role of packing effects.\par
\begin{figure}

  \centering	
  \includegraphics[width=0.475\textwidth]{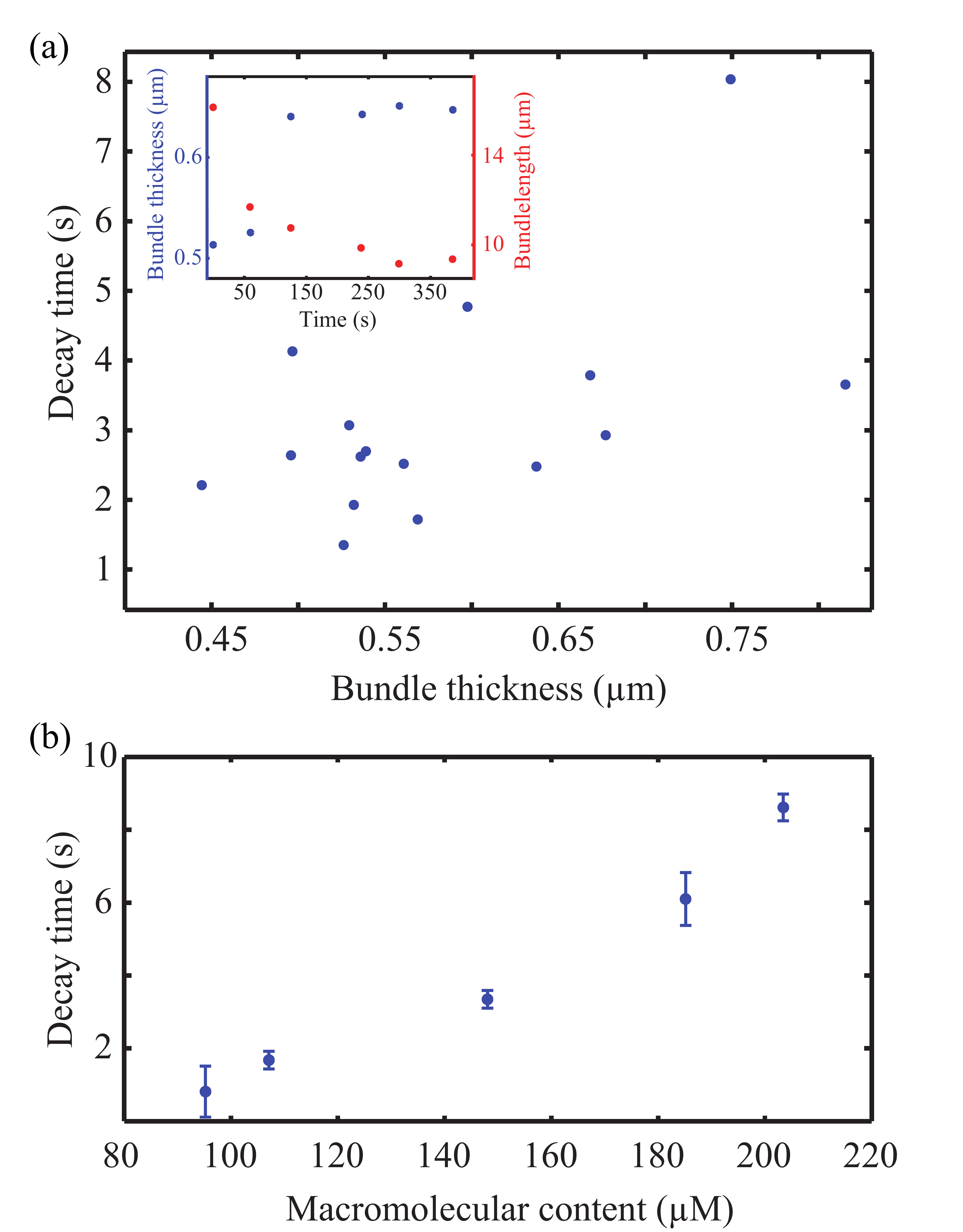}
    \caption{(Color online) (a) The bundle thickness has no detectable scaling behavior within the limits of our measurements. (Inset) In the course of a contraction process (red dots) a thickening of the contracting bundle (blue dots) can be observed . (b) With an increasing molecular content of the depletion agent an increase of decay times was observed. Thus, a more viscous medium yields a slower bundle contraction due to higher friction.}
	\label{fig:thicknessviscosity}
	
\end{figure}

Evaluations of the dynamic behavior of contractions enable us to test influences of the contraction process on the bundle itself. We monitored the bundle thickness over time (Fig. \ref{fig:thicknessviscosity} (a)) and found that a bundle becomes thicker during the contraction since overlapping filaments are driving the system to a shorter structure. These measurements aimed at comparing different bundle thicknesses to contractile dynamics. Bundle widths were evaluated via a Radon transform along the bundle backbone \cite{Zhang.2007}. Some of these measurements were discarded since these evaluations showed a dependency on input parameters. Presented data, however, were consistent and evaluations were done with the same parameters. The full width at half maximum of the intensity profiles was chosen to compare thicknesses of different bundles. Due to the direct correlation to interacting filament pairs we expect an influence of the overall bundle thickness on exerted forces and kinetics during contractions of different bundles. However, we found no apparent correlation within the limits of our measurements  (Fig. \ref{fig:thicknessviscosity} (a)). This data is in good agreement with consistent decay times for accelerating events, where a bundle thickening has seemingly a minor influence. Possibly, effects due to thicker bundles are superimposed by unavoidable viscosity variations for differing experiments. These variations correspond to different macromolecular contents of the depletion agent influencing the contraction process. We observed that a higher macromolecular content inherently slows down contractions due to higher friction forces and shows approximately a linear dependency with respect to the decay times of contraction processes (Fig. \ref{fig:thicknessviscosity} (b)). For this investigation the viscosity was measured by optical tweezers as described in \cite{TolicNorrelykke.2006} with a bead in vicinity to the bundle. Viscosity values were directly translated into macromolecular contents with data sheets provided by Sigma-Aldrich.\par
\begin{figure}
  \centering	
  \includegraphics[width=0.475\textwidth]{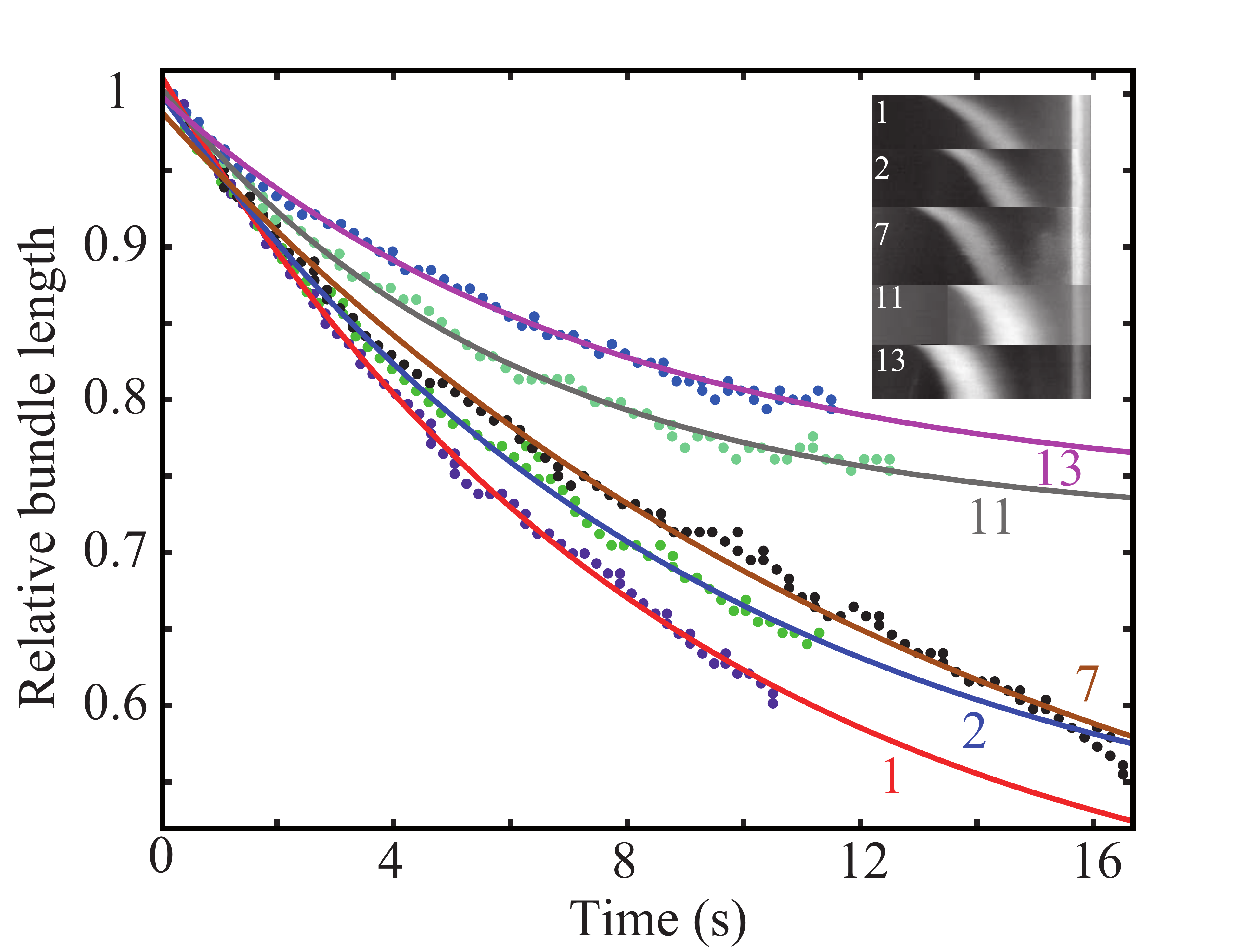}
    \caption{(Color online) Contractions show a decaying behavior when one bundle is deformed multiple times consecutively (Movie S 2). Dynamic behavior becomes slower with every expansion and contraction event and the bundle relaxes to other baselines. Numbers refer to the specific contraction process. Image series were evaluated in form of a kymograph. The relative bundle length describes the actual bundle length normalized by the maximal outstretched configuration in the according experiment.}

	\label{fig:multiplecontractions}
	
\end{figure}

As a further test of influences of the contraction process to the bundle, we investigated the persistence of a bundle's contraction behavior. In that course we deformed a single bundle multiple times and recorded its contraction behavior. For a better experimental realization bundles attached solely to one bead were probed. Our experiments revealed a degenerating effect after consecutive expansions and contractions. As displayed in Fig. \ref{fig:multiplecontractions}, later contractions display lower maximal velocities and reach a higher baseline representing an increased relaxed bundle length. We attribute this fact to potential filament annealing (two filaments concatenate) yielding a change in the energy balance \cite{Andrianantoandro.2001}. For further contractions these merged filaments would have to buckle and thus hinder the overall contraction process.\par
In conclusion, we developed an optical tweezers based technique to investigate the contractile behavior induced by depletion forces \cite{Hosek.2004} of a multi-filament actin bundle. In comparison to previous theoretical as well as experimental studies \cite{Kinoshita.2004, Li.2005, Galanis.2010, Hilitski.2014} we found a fundamentally different, dynamic behavior. These earlier studies described that a relative, axial sliding of single rod-like filaments induced by depletion forces leads to a linear force exertion.  Dynamics of contractions would then proceed with a constant velocity, at odds with our findings of an exponentially decreasing velocity. We are able to describe this behavior as an emergent phenomenon of rod-like colloids in an actin bundle when taking pairwise interactions to a multi-filament scale. To further understand the results of our experiments we model the bundle as a simple two-dimensional arrangement of N-1 laterally stacked pairs of rigid filaments. The arising harmonic potential and accordingly the exponential force decay were verified by simulations. To measure absolute force values, different techniques have to be applied like in \cite{Hilitski.2014}, but these methods hardly allow evaluations of dynamics. \par
Molecular crowding effects represent a fundamental physical interaction, which cannot be switched off even in active systems such as cells. The cytoplasm itself is a dense environment filled with macromolecules \cite{Ellis.2001}. In the experiments presented here, however, the amount of macromolecules is well below the macromolecular content of a cell emphasizing the biological relevance (see supplement). Additionally, kinetics and force generation are in a regime of active processes but without the need to convert chemical energy into mechanical work \cite{Lodish.op.2000,Clemen.2005,Finer.1994,Kron.1986,Scholey.1993}.\par
\begin{acknowledgments}
We like to thank Jessica Lorenz helping to set up the SDS-PAGE and David M. Smith as well as Tina Händler for proofreading this manuscript and helpful discussions. This work was supported by the graduate school “Building with Molecules and Nano-Objects” (BuildMoNa – GSC 185/1) and the DFG Forschergruppe (FOR 877). CH acknowledges the support of the German Science Foundation (DFG) via the Emmy Noether fellowship He 6322/1-1 as well as via the collaborative research center SFB 937, project A16. \par
\end{acknowledgments}

\bibliography{bibliography}

\end{document}


\preprint{APS/123-QED}

\title{Supplementary Material\\
Motor-free actin bundle contractility driven by molecular crowding}

\author{Jörg Schnauß$^1$}
\author{Tom Golde$^1$}
\author{Carsten Schuldt$^1$}
\author{B. U. Sebastian Schmidt$^1$}
\author{Martin Glaser$^1$}
\author{Dan Strehle$^1$}
\author{Josef A. Käs$^1$}
\author{Claus Heussinger$^2$}
\affiliation{1 Institute for Experimental Physics I, University of Leipzig , Linnéstraße 5, 04103 Leipzig , Germany}
\affiliation{2 Institute for Theoretical Physics, Georg-August University of Göttingen, Friedrich-Hund Platz 1, 37077 Göttingen, Germany}

\maketitle
\section{I Biochemical preparation and experimental methods}

\begin{figure}[b]
  \centering	
	\vspace{-1cm}
  \includegraphics[width=0.4\textwidth]{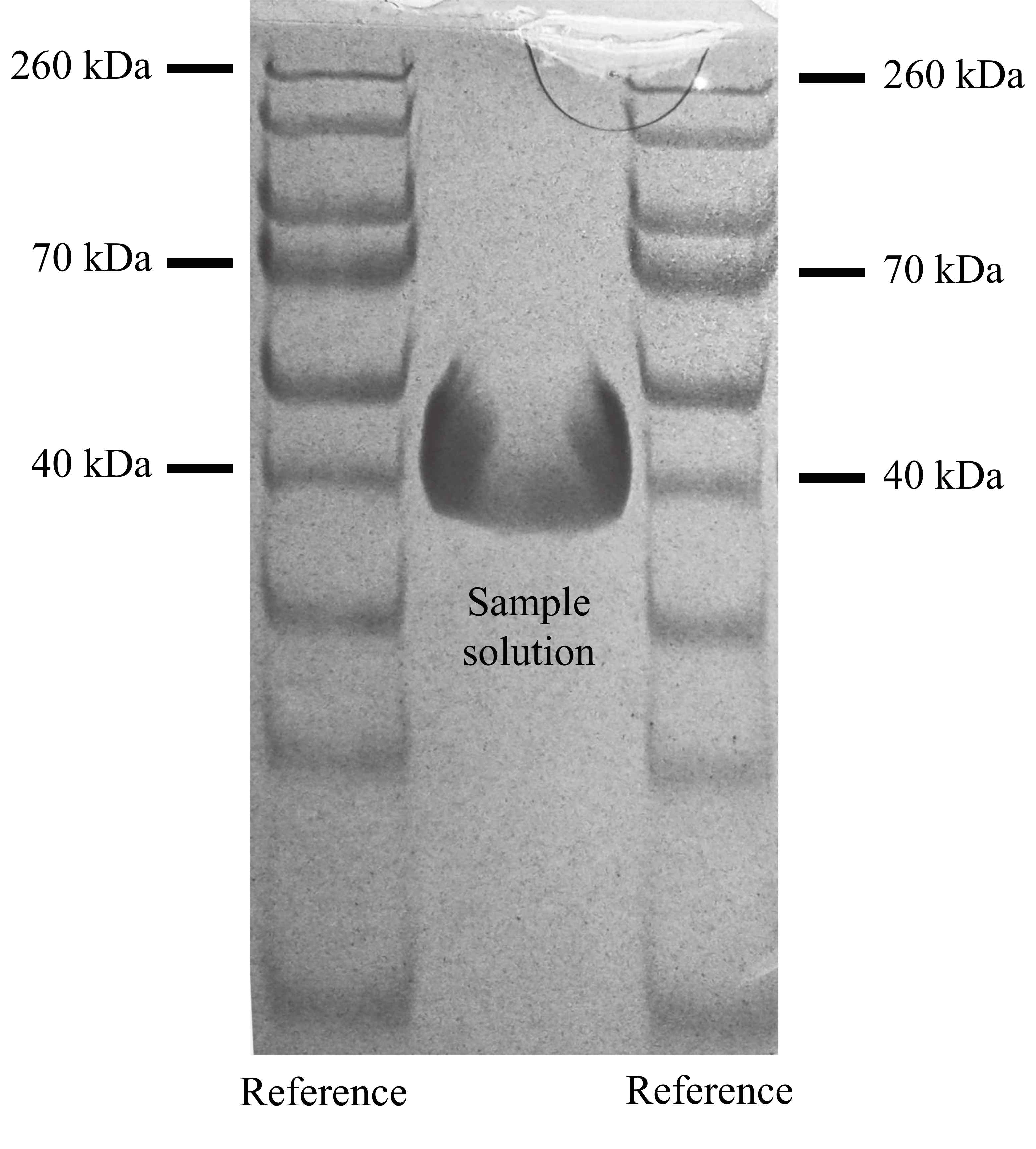}
    \caption{SDS-PAGE was used to show the absence of any myosin motors or other accessory proteins in solution. Only the actin signal at \SI{42}{\kilo\dalton} was observed verifying the purity of the solutuion.}
	\label{fig:SDS}
\end{figure}

G-actin was prepared from rabbit muscle as described previously \cite{Gentry.2009}. This procedure involved size-exclusion chromatography to ensure the purity of the actin and the absence of other proteins. This purity was verified by SDS-PAGE (Fig. \ref{fig:SDS}). We observed the actin signal at \SI{42}{\kilo\dalton} but no other signals, which would indicate additional components. To ensure a complete independence of myosins, we verified the persistence of the effect under ADP conditions rendering active processes impossible. The presence of any crosslinking proteins would have inhibited the initial experimental step since sliding motions would have been suppressed.\\
Monomeric actin was polymerized and labelled by adding $1/10$ volume fraction of 10 times concentrated F-buffer (\SI{1.0}{\Molar} KCl, \SI{10}{\milli\Molar} MgCl2, \SI{50}{\milli\Molar} HEPES, \SI{2.0}{\milli\Molar} ATP or ADP, \SI{10}{\milli\Molar} DTT) and Phalloidin–Tetramethylrhodamine~B isothiocyanate (Phalloidin-TRITC - Sigma-Aldrich Co.) at a concentration of \SI{5}{\micro\Molar}. 
Biotinylated actin (\SI{5}{\micro\Molar} - Cytoskeleton Inc.) was added to the solution to decorate filament ends after polymerization (ratio of 3 to 50). Further actin preparation was done as described above and polymerization was induced by ADP-F-Buffer conditions.\\
For experiments under ADP conditions, all buffers were prepared with ADP instead of ATP. To exchange the former G-Buffer (\SI{5}{\milli\Molar} Tris-HCl, \SI{0.1}{\milli\Molar} CaCl2, \SI{0.2}{\milli\Molar} ATP, \SI{1.0}{\milli\Molar} DTT, \SI{0.01}{\percent} NaN3), G-actin was polymerized under ADP-F-Buffer conditions and centrifuged at 100,000 g for 3.5 h at 4°C. The pellet was resuspended in ADP-G-Buffer (\SI{5}{\milli\Molar} Tris-HCl, \SI{0.1}{\milli\Molar} CaCl2, \SI{0.2}{\milli\Molar} ADP, \SI{1.0}{\milli\Molar} DTT, \SI{0.01}{\percent} NaN3) and dialyzed against ADP-G-Buffer overnight.\\
\SI{2}{\micro\meter} fluorescent streptavidin beads (Streptavidin Fluoresbrite® YG Microspheres - Polysciences Inc.) were used to allow binding of the prepared, labelled actin via biotin-streptavidin bonds. Beads were captured by optical tweezers enabling contact-free manipulations and measurements of dynamics within the system.\\
Depletion forces were induced by adding methyl cellulose (400 cP (\SI{2}{\percent} in aqueous solution) - Sigma-Aldrich Co.) to arrange actin filaments into bundles without the need of accessory proteins \cite{Strehle.2011}. The final solution contained \SI{0.2}{\micro\Molar} TRITC-labelled, biotinylated actin, glucose/glucose oxidase as antiphotobleaching agent, and \SI{1.6}{\percent} methyl cellulose in F-buffer conditions. This solution was deposited into a sample chamber as described previously \cite{Strehle.2011}.\\
Fluorescence was induced by a mercury vapor lamp and a N2.1 filter cube transmitting the appropriate green light (Leica 11513882, excitation filter from 515 to \SI{560}{\nano\meter}) on the sample. In this spectrum actin bundles as well as beads were visualized.\\
To manipulate beads and actin structures an epi-fluorescence Leica DM IRB microscope equipped with a 100×oil-immersion objective (Leica 11506168) and a custom-built optical tweezers setup was used (similar to the setup described by Koch et al. \cite{Koch.2004} equipped with a Manlight ML3-CW-P-OEM-OTS laser). Observations and image acquisitions were realized by a Hamamatsu Orca ER digital CCD camera (Hamamatsu Photonics). All components of the setup were controlled and integrated by a self-written LabVIEW (National Instruments) program. Experiments were observed via image sequences with a typical frame rate of \SI{20}{\hertz}.\\

\section{II Forces and velocities}

\begin{figure}[h]
  \centering	
  \includegraphics[width=0.4\textwidth]{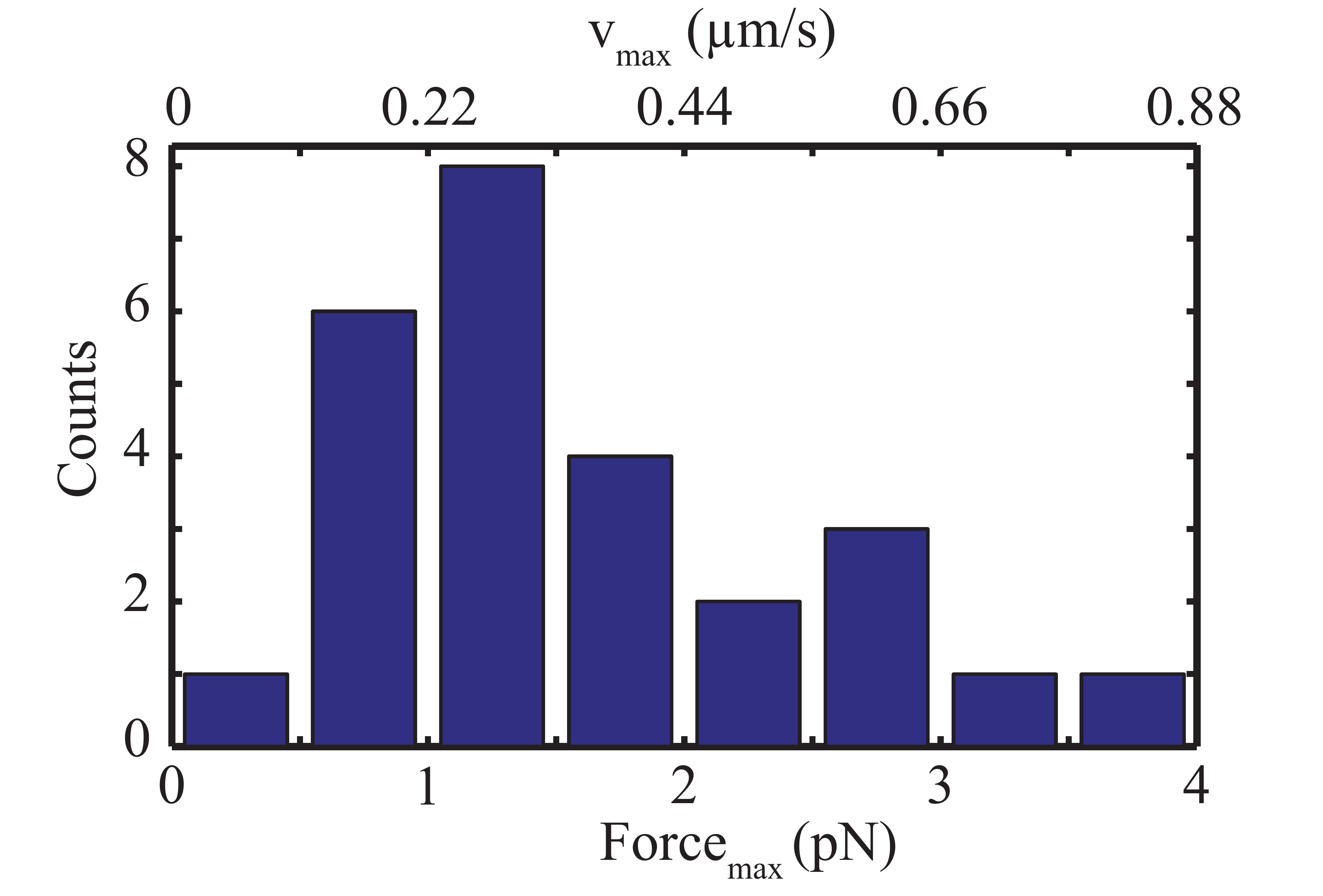}
    \caption{Maximal velocities (typically ranging from 0.10 to \SI{0.65}{\micro\meter \per \second}) and forces (typically ranging from 0.5 to \SI{3.0}{\pico\newton}) of contractile bundles attached to two beads.}
	\label{fig:forces}
\end{figure}

Exponential decay functions describing the bundle length over time during the contraction process can be used to further evaluate these processes. The fitting function can be analytically differentiated yielding the velocity of the contraction. Resulting maximal velocities of the contractions typically ranged from  0.10 to \SI{0.65}{\micro\meter \per \second} (Fig. \ref{fig:forces}). If two beads were attached to a bundle, the bead's geometry was used to determine contractile forces. The released bead was dragged through the solution by the contraction. Forces, necessary to displace a bead in a viscous medium, can be approximated by Stokes’ law ($\mathrm{F=6 \pi \eta rv}$, with $\eta$ being the viscosity, r the radius of the bead, and v the velocity). This approach, however, systematically underestimates forces since longitudinal friction at the bundle surface is neglected. For this evaluation the viscosity of the surrounding medium was measured with microrheological methods. Measurements yielded a viscosity of $\eta$ = 0.24 $\pm$ \SI{0.03}{\pascal\second}. Our data have shown maximal forces typically range from 0.5 to \SI{3.0}{\pico\newton} (Fig. \ref{fig:forces}). 
A spread of maximal forces is evident since bundles are very heterogeneous structures and uniform starting conditions are not feasible. Bundle thicknesses, for instance, naturally vary and the number of filaments within a bundled structure can only be approximated \cite{Strehle.2011}. The amount of free filament ends can vary drastically from bundle to bundle. Additionally, maximal velocities as well as forces depend on the extension of the bundle during the stretching process, which is limited by the strength of the optical tweezers setup.\\
Due to the exponential length decay, generated forces converge to zero when bundles are fully contracted approaching the thermal noise regime. Forces due to thermal noise, however, are not directed. Contractile forces in our experiments are directed and thus bundles still relax even in this low force regime until reaching their energetic minimum.\\

\section{III Simulation of the cross-over to a multi-filament system}

\begin{figure}[h]
  \centering	
  \includegraphics[width=0.4\textwidth]{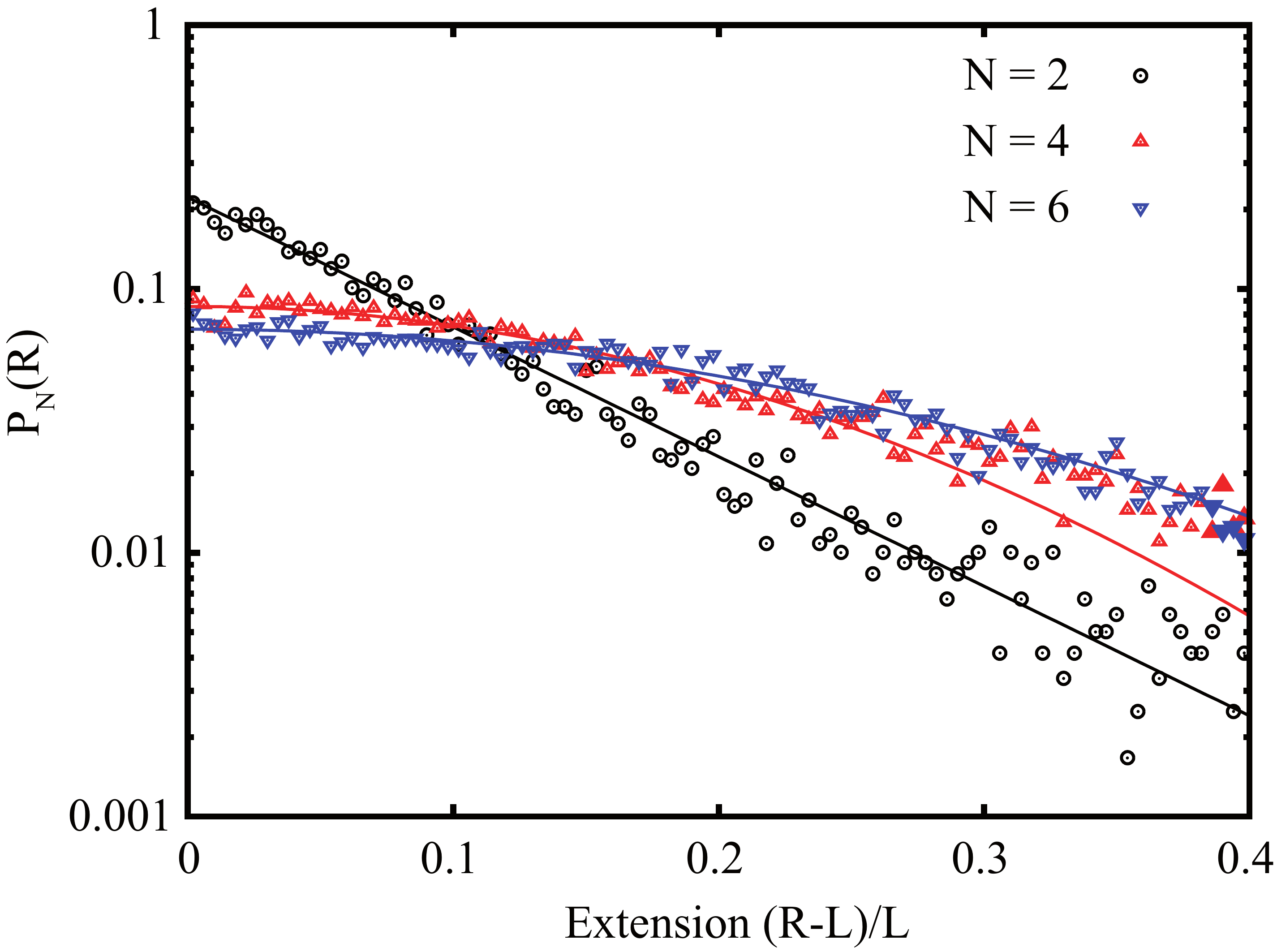}
    \caption{Probability distributions P(R) for the extensions (end-to-end distance) of bundles of N = 2, 4, 6 filaments are displayed. Points are direct results of the simulation and lines are corresponding fits. For a bundle of two filaments the probability distribution follows an exponential distribution as given by the Boltzmann weight. Probability distributions of a bundle with more filaments follow a Gaussian distribution. This transition is illustrated by simulating N = 2, 4, 6 filament bundles. Already bundles formed by six filaments yield a Gaussian probability distribution corresponding to a harmonic potential.}
	\label{fig:simulation}
\end{figure}

The analytical model describes the free energy of the system as a function of the end-to-end distance R quickly reaching an asymptotic form already for a few filaments. We are able to confirm this behavior by molecular dynamics simulations, where the depletant is modeled explicitly via soft spheres. Filaments are modeled as rigid rods and excluded volume effects via shifted and truncated Lennard-Jones interactions for filament-filament and filament-depletant pairs. The ratio between filament and depletant diameter is chosen to be five. The filament length is 50 in units of the depletant diameter, and the volume fraction of the depletant is taken to be 0.5.\\
In Fig. \ref{fig:simulation} the probability distribution for the offset between first and last filament in the bundle (no force, f = 0) is shown. For an N = 2 bundle an exponential distribution arises as given by the Boltzmann weight (exp(-$\mathrm{\beta u \left| x \right|}$)). Upon addition of more filaments to the bundle a cross-over to a Gaussian distribution can be observed. This distribution corresponds to a harmonic energy profile.\\
Distributions of more filaments (e.g. N = 4, 6) can be well described by Gaussian fits, in particular for small offsets. Furthermore, the fit for N = 6 resembles also the behavior for larger offsets as expected from the theory. The relative width $\sigma$ of the Gaussian fits for the two cases of N = 4,6 filament bundles yield $\mathrm{\sigma_6 / \sigma_4 \approx \sqrt{6/4}}$. This behavior is as expected from the N-dependence of the spring constant k $\sim$ 1/N and $\mathrm{\sigma \sim \sqrt{1/k}}$.

\section{IV Biological relevance}

Especially for forces in the piconewton range, the passive process we found cannot be ignored. These contributions minimize the free energy of the bundle if filaments interact in an attractive fashion. Our experiments only rely on filament-filament interactions induced by the environment and no other proteins besides actin are involved. In addition, contractions do not require a specific orientation of actin filaments concerning their plus and minus ends. Dynamics of the energetically driven processes, however, are slow in comparison to velocity ranges of some myosins. Myosin II motors, for instance, have been reported to move up to \SI{4.5}{\micro\meter \per \second} \cite{Lodish.op.2000, Carvalho.2013, Finer.1994, Clemen.2005}. Myosin types moving actively in a lower velocity range have also been described \cite{Scholey.1993}. Besides dynamic differences, myosin activity is switchable while filament interactions cannot be influenced easily \cite{Stachowiak.2012}.\\
Although dynamics of the energy minimizing processes are slow, they can be employed to contribute to contractions. The passive contractility process we found may act independently of molecular motors but may also act in concert with motor activity. Cellular actin systems, for instance, are prestressed by myosin minifilaments \cite{Gardel.2006}. These motors maintain tensile strains of the bundle and the overall cellular strain can be preserved. However, myosin motors bind transiently and minifilaments can be very small containing only three to four myosins. Thus, there are states when all molecular motors are detached from the bundle. Due to its prestress a bundle would extend its length decreasing its tension. In extreme cases bundles would even fully dissolve. Nevertheless prestressed bundles remain intact since filament-filament interactions antagonize bundle disintegration. Tensile strains of bundles and consequently of whole cells can be sustained.\\
On a cellular scale this mechanism may play a role in the retraction phase of filopodia, which are among the most molecularly crowded structures in cells. Actin filament bundles stretching from the tip to the base of filopodia are highly contractile although they are not directly contracted by the myosin motors located at the basal interface to the interior actin cortex \cite{Romero.2012}. This is further supported by the fact that bundled actin filaments within filopodia are highly parallel in polarity due to their polymerization from the tip, thereby rendering active myosin contractions implausible. In contrast, the fundamental mechanism underlying contraction presented here is entirely independent of the orientation of interacting filament pairs. Stall forces for retracting filopodia are on the same order of magnitude as the crowding-induced contractile forces measured in our experiments \cite{Romero.2012}. Their kinetics, however, seem to be slower \cite{Bornschlogl.2013}, which might be attributed to crosslinkers such as fascin or fimbrin regulating the depletion force effect. Fascin itself might inhibit these depletion driven contractions due to its properties as a relatively stable crosslinker on the time scale of retraction events \cite{Vignjevic.2006}. However, fascin’s binding affinity can be influenced by phosphorylation, which drastically reduces its actin binding affinity \cite{Yamakita.1996,Ono.1997}. This mechanism, which itself is regulated by interactions between the surface of the cell and its surrounding environment, might act as a trigger to initiate bundle, and thereby filopodia contraction via crowding-induced forces.\\
The \textit{in vivo} situation in cellular systems, however, is far more complex with many interacting components and interwoven functions. Examinations of a minimal system of actin filaments in a crowded environment initially allowed us to investigate distinct biophysical properties, which are hardly accessible in \textit{in vivo} systems. However, extensive \textit{in vivo} studies are necessary to determine processes and magnitudes of entropic forces induced by a crowded environment in active systems such as cells.\\
In conclusion, these contractions are based on fundamental physical laws and are very robust. Entropic crowding effects can generate forces in the piconewton range which are relevant in biological systems since they are comparable to myosin activity. This statement is emphasized by the macromolecular content employed for our experiments, which is well below the macromolecular content of cells. Furthermore, crowding effects cannot be switched off and bundle contractions always appear if filaments can slide against each other. Thus, depletion forces and accordingly these contractile actin structures should appear in cellular systems as well. They are independent of any active structural arrangements and parameters such as filament orientation.\\

\bibliography{bibliography}